\documentclass[twocolumn,preprintnumbers,amsmath,amssymb,nobibnotes,nofootinbib,superscriptaddress]{revtex4}
\usepackage{graphicx}
\usepackage{dcolumn}
\usepackage[usenames,dvipsnames]{xcolor}
\RequirePackage[colorlinks=true
,urlcolor=blue
,anchorcolor=blue
,citecolor=blue
,filecolor=blue
,linkcolor=blue
,menucolor=blue
,pagecolor=blue
,linktocpage=true
,pdfproducer=medialab
]{hyperref}
\usepackage{amsmath}
\usepackage{mathrsfs}
\usepackage{slashed}
\usepackage{cancel}
\usepackage{feynmp}

\newcommand{\LL}{\mathscr{L}}
\newcommand{\cG}{\mathscr{G}}

\def\cF{{\cal F}}

\def\cG{{\cal G}}
\def\cGSM{{\cal G}_{SM}}

\def\cM{{\cal M}}
\def\cN{{\cal N}}
\def\cO{{\cal O}}

\def\cP{{\cal P}}
\def\cR{{\cal R}}

\def\cW{{\cal W}}

\def\Tr{{\rm Tr}}

\def\be{\begin{equation}}
\def\ee{\end{equation}}
\def\beq{\begin{equation}}
\def\eeq{\end{equation}}
\def\bc{\begin{center}}
\def\ec{\end{center}}
\def\bea{\begin{eqnarray}}
\def\eea{\end{eqnarray}}

\def\nt{\noindent}
\newcommand{\hc}{\mathrm{h.c.}}

\newcommand{\derp}{\partial}

\newcommand{\UH}{\mathbf{U}}
\newcommand{\UHL}{\mathbf{U}_L}
\newcommand{\UHR}{\mathbf{U}_R}
\newcommand{\UHLR}{\mathbf{U}_{L(R)}}

\newcommand{\TL}{\mathbf{T}}
\newcommand{\VL}{\mathbf{V}}
\newcommand{\DL}{D}

\newcommand{\VLLmuu}{\mathbf{V}^\mu_L}
\newcommand{\VLLmud}{\mathbf{V}_{\mu,\,L}}

\newcommand{\VLRmuu}{\mathbf{V}^\mu_R}
\newcommand{\VLRmud}{\mathbf{V}_{\mu,\,R}}

\newcommand{\VLLmuut}{\widetilde{\mathbf{V}}^\mu_L}

\newcommand{\VLRmudt}{\widetilde{\mathbf{V}}_{\mu,\,R}}

\newcommand{\BBu}{B^{\mu\nu}}
\newcommand{\BBd}{B_{\mu\nu}}

\newcommand{\WWLut}{\widetilde{W}^{\mu\nu}_L}

\newcommand{\WWRdt}{\widetilde{W}_{\mu\nu,\,R}}

\newcommand{\TLchi}{\mathbf{T}_\chi}
\newcommand{\VLchimuu}{\mathbf{V}^\mu_\chi}
\newcommand{\VLchimud}{\mathbf{V}_{\mu,\,\chi}}

\newcommand{\WWchiu}{W^{\mu\nu}_\chi}

\newcommand{\WWchiut}{\widetilde{W}^{\mu\nu}_\chi}

\newcommand{\TLLR}{\mathbf{T}_{L(R)}}
\newcommand{\VLLRmuu}{\mathbf{V}^\mu_{L(R)}}

\newcommand{\TLLt}{\widetilde{\mathbf{T}}_{L}}
\newcommand{\TLRt}{\widetilde{\mathbf{T}}_{R}}

\newcommand{\fL}{f_L}
\newcommand{\fR}{f_R}
\newcommand{\fLR}{f_{L(R)}}

\newcommand{\fchi}{f_\chi}
\newcommand{\gL}{g_L}
\newcommand{\gR}{g_R}

\newcommand{\gchi}{g_\chi}

\begin{document}
%
%

\title{Diboson excess and $Z'$--predictions via left-right non--linear Higgs}

\author{Jing Shu}
\email{jshu@itp.ac.cn}
\affiliation{State Key Laboratory of Theoretical Physics and Kavli Institute for Theoretical Physics China (KITPC)\\
Institute of Theoretical Physics, Chinese Academy of Sciences, Beijing 100190, P. R. China}

\author{Juan Yepes}
\email{juyepes@itp.ac.cn}
\affiliation{State Key Laboratory of Theoretical Physics and Kavli Institute for Theoretical Physics China (KITPC)\\
Institute of Theoretical Physics, Chinese Academy of Sciences, Beijing 100190, P. R. China}

\begin{abstract}

The excess events reported by the ATLAS Collaboration in the $WZ$--final state, and by the CMS Collaboration in the $e^+\!e^- jj$, $Wh$ and $jj$--final states, may be induced by the decays of a heavy boson $W'$ in the 1.8--2 TeV mass range, here modelled via the larger local group $SU(2)_L\times SU(2)_R\times U(1)_{B-L}$ in a non--linear dynamical Higgs scenario. The $W'$--production cross section at the 13 TeV LHC is around 700--1200 fb. This framework also predicts a heavy $Z'$ boson with a mass of 2.5--4 TeV, and some decay channels testable in the LHC Run II. We determine the cross section times branching fractions for the dijet, dilepton and top--pair $Z'$--decay channels at the 13 TeV LHC around 2.3, 7.1, 70.2 fb respectively, for $M_{Z'}= 2.5$ TeV, while one/two orders of magnitude smaller for the dijet/dilepton and top--pair modes at $M_{Z'}= 4$ TeV. Non-zero contributions from the effective operators, and the underlying Higgs sector of the model, will induce sizeable enhancement in the $W^+W^-$ and $Z h$--final states that could be probed in the future LHC Run II.

\end{abstract}
\maketitle


\section{Introduction}

\nt Tantalizing deviations from the SM predictions have been recently reported by the ATLAS and CMS Collaborations around invariant mass  of 1.8--2 TeV, and are claiming for:

\begin{itemize}

\item[\bf a)] $3.4\sigma$  local ($2.5\sigma$ global) excess in the ATLAS search~\cite{ATLAS1} (CMS reports a slight excess at the same mass~\cite{CMS-VV}) for a heavy resonance $W^\prime$ decaying as $W^\prime \to WZ \to JJ$, where $J$ stands for two colinear jets from a $W$ or $Z$--boosted decay;

\item[\bf b)] $2.8\sigma$ excess in the CMS search~\cite{CMS}  for a heavy right handed boson $W^\prime$ decaying  into an electron and a right handed neutrino $N$, as $W^\prime \to N\,e \to eejj$; 

\item[\bf c)] $2.2\sigma$ excess in the CMS search~\cite{CMS1} for $W^\prime\to W h$, with a highly boosted SM Higgs boson $h$ decaying as $h \to b\bar{b}$ and $W\to \ell \nu$ (with $\ell=e,\mu$);

\item[\bf d)] $2.1\sigma$ excess in the CMS dijet search~\cite{CMS2}.

\end{itemize}

\nt In spite of requiring more statistics at the LHC Run II to shed light on their real origin, and being not significant  enough to point out BSM new phenomenon, it is worthwhile to explore which features are motivated by such deviations in a given theoretical framework. In this regard, many models and scenarios have been proposed. Among them, the left--right EW symmetric model, based on the gauge group $\cG=SU(2)_L\times SU(2)_R\times U(1)_{B-L}$~\cite{LRSM1,LRSM2}, seems to address properly the observed excesses in all the mentioned decay channels. Indeed, the $WZ$ excess (item \textbf{a}) and $W h$ excess (item \textbf{c}) can be tackled~\cite{Dobrescu:2015qna, Gao:2015irw, Brehmer:2015cia} via $W^\prime\to WZ,~W h$, as the implied couplings arise naturally in these models (see~\cite{other} for some alternative explanations of the diboson excess). The $eejj$ excess (item \textbf{b}) can be understood~\cite{Dobrescu:2015qna, Deppisch:2014qpa, Fowlie:2014iua, Gluza:2015goa} through the process $pp\to W^\prime\to N\,e\to eejj$~\cite{KS}, and for a charged gauge boson mass $M_{W^\prime}\sim 2$ TeV, with $g_R<g_L$ at the TeV-scale~\cite{Dobrescu:2015qna}. Finally, the dijet excess (item \textbf{d}) may simply be yielded by $W^\prime\to jj$.  

The observed excess events are interpreted in this work as being induced by the decays of a heavy boson $W'$ with a mass range 1.8--2 TeV, where the underlying framework relies in a non--linearly realized left--right model coupled to a light Higgs particle. Calling for the larger local group $\cG=SU(2)_L\times SU(2)_R\times U(1)_{B-L}$ in an electroweak non--linear $\sigma$--model, the Goldstone bosons are parametrized as customarily  via the dimensionless unitary matrices $\UHL(x)$ and $\UHR(x)$ for the symmetry group $SU(2)_L\times SU(2)_R$, and defined as
\beq
\UHLR\,(x)=e^{i\,\tau_a\,\pi^a_{L(R)}(x)/\fLR}\, , 
\label{Goldstone-matrices}
\eeq

\nt with $\pi^a_{L(R)}(x)$ the corresponding GB fields suppressed by their associated non--linear sigma model scale $\fLR$. In addition, this non--linear effective set--up is coupled a posteriori to a Higgs scalar singlet $h$ through powers of $h/\fL$~\cite{Georgi:1984af}, via the generic light Higgs polynomial functions $\cF(h)$~\cite{Alonso:2012px}
\beq
\cF_i(h)\equiv1+2\,{a}_i\,\frac{h}{\fL}+{b}_i\,\frac{h^2}{\fL^2}+\cO\Big(\frac{h^3}{\fL^3}\Big)\,.
\label{F}
\eeq

\nt This work is split into: Sect.~\ref{EffectiveLagrangian} describes the EW effective Lagrangian following the light dynamical Higgs picture in~\cite{Alonso:2012px,Brivio:2013pma,Gavela:2014vra,Alonso:2012pz,
Yepes:2015zoa} (see also Ref.~\cite{Buchalla:2013rka,Buchalla:2012qq,Buchalla:2013eza} and~\cite{Brivio:2015kia} for a Higgs portal to scalar dark matter in non-linear EW approaches), focused only in the CP--conserving bosonic operators\footnote{See ~\cite{Cvetic:1988ey,Alonso:2012jc,Alonso:2012pz,Buchalla:2013rka} for non--linear analysis including fermions.}. The mixing effects for the gauge masses triggered by the LRH operators and the corresponding gauge physical masses are also analysed there. Sect.~\ref{Diboson-excess-W'-production} analyses the $W'$--production and the constraints on the parameter space of our scenario entailed by the reported excesses in the $WZ$ and $W h$--final states. Sect.~\ref{Z'--predictions} explores the prediction of a heavy boson $Z'$ in the model, its possible mass range and the implied dijet, dilepton and top--pair decay channels. The less dominant decays $Z^\prime \to \{W^+W^-,\, Z h\}$, and the 
sizeable enhancement they can suffer by the physical impact of non-zero contribution from the effective non--linear operators is also analysed. Finally, Sect.~\ref{Conclusions} summarizes the main results.

\section{Effective Lagrangian}
\label{EffectiveLagrangian}

\nt The NP departures with respect to the SM Lagrangian $\LL_0$ and will be encoded in this work through the effective Lagrangian 
\be
\begin{aligned}
\LL_\text{chiral} = \LL_0\,+\,\LL_{0,R}
\,+\,\Delta \LL_{\text{CP}}\,+\,\Delta\LL_{\text{CP},LR}\,.\\[-4mm]
\label{Lchiral}
\end{aligned}
\ee 

\nt The first three pieces in $\LL_\text{chiral}$ read as
\be
\begin{aligned}
&\LL_0 =\\
& -\dfrac{1}{4}\,\BBd\,\BBu-\dfrac{1}{4}\,W^a_{\mu\nu,\,L}\,W^{\mu\nu,\,a}_L-
\dfrac{1}{4}\,G^a_{\mu\nu}\,G^{\mu\nu,\,a}\,+\\[2mm]
&+\frac{1}{2} (\derp_\mu h)(\derp^\mu h) - V (h)-\dfrac{\fL^2}{4}\Tr\Big(\VLLmuu\VLLmud\Big)\left(1+\frac{h}{\fL}\right)^2\,+\\[2mm]
&+i\bar{q}_L\slashed{D}q_L+i\bar{l}_L\slashed{D}l_L\,,
\end{aligned}
\label{LLO}
\ee

\be
\begin{aligned}
&\LL_{0,R}=\\
&-\dfrac{1}{4}\,W^a_{\mu\nu,\,R}\,W^{\mu\nu,\,a}_R\,-\,\frac{\fR^2}{4}\,\Tr\Big(\VLRmuu\,\VLRmud\Big)\left(1+\frac{h}{\fL}\right)^2\,+\\[2mm]
&+i\bar{q}_R\slashed{D}q_R+i\bar{l}_R\slashed{D}l_R\,,
\end{aligned}
\label{LLO-Right}
\ee

\nt where the adjoints $SU(2)_{L(R)}$--covariant vectorial $\VLLRmuu$ and the covariant scalar $\TLLR$ are defined as
\be
\VLchimuu \equiv \left(\DL^\mu \UH_\chi\right)\,\UH^\dagger_\chi\,, \quad
\TLchi \equiv \UH_\chi\,\tau_3\,\UH^\dagger_\chi\,,
\label{EFT-building-blocks}
\ee
\nt with $\chi=L,R$ and the corresponding covariant derivative for both of the Goldstone matrices $\UHLR(x)$ introduced as
\be
\DL^\mu \UH_\chi \equiv \derp^\mu \UH_\chi \, + \,\frac{i}{2}\,\gchi\,W^{\mu,a}_\chi\,\tau^a_\chi\,\UH_\chi - \frac{i}{2}\,g'\,B^\mu\,\UH_\chi\,\tau^3 \,
\label{Covariant-derivatives}
\ee
\nt where the $SU(2)_L$, $SU(2)_R$ and $U(1)_{B-L}$ gauge fields are denoted by $W^{a\mu}_L$, $W^{a\mu}_R$ and $B^\mu$ correspondingly, and the associated gauge couplings $\gL$, $\gR$ and $g'$ respectively. The scale factor of $\Tr\left(\VLLmuu\,\VLLmud\right)$ entails GB--kinetic terms canonically normalized, in agreement with the $\UHL$--definition in~\eqref{Goldstone-matrices}. The corresponding $SU(2)_R$--counterparts for the strength gauge kinetic term and the custodial conserving operator at the Lagrangian $\LL_0$ are parametrized by $\LL_{0,R}$ in~\eqref{LLO-Right}, entailing thus an additional scale $\fR$ that encodes the new high energy scale effects introduced in the scenario once the SM local symmetry group $\cGSM$ is extended to $\cG$.  
The associated fermion kinetic terms are described by the 3rd and 2nd lines in~\eqref{LLO}-\eqref{LLO-Right} respectively, with the quark and lepton doublets $q^i$ and $l^i$ ($i$ stands for fermion generations) defined as
\be
\hspace*{-1mm}
\begin{aligned}
q^i_{L}&\!=\left(\begin{array}{c}u^i_{L}\\d^i_{L}\end{array}\right) \sim (2,1,1/6),\,\quad 
&l^i_{L}&\!=\left(\begin{array}{c}\nu^i_{L}\\e^i_{L}\end{array}\right) \sim (2,1,-1/2), \\ \\
q^i_{R}&\!=\left(\begin{array}{c}u^i_{R}\\d^i_{R}\end{array}\right) \sim (1,2,1/6),\, \quad
&l^i_{R}&\!=\left(\begin{array}{c}N^i_{R}\\e^i_{R}\end{array}\right) \sim (1,2,-1/2),
\end{aligned}
\ee

\nt where it have been specified the transformation properties under 
the group $\cG$ corresponding to the usual fermion representation for the left-right models. The right-handed neutrinos $N_R^i$  acquire masses at the TeV scale through the mechanism of Ref.~\cite{Coloma:2015una}. The scalar sector includes in general an $SU(2)_R$ doublet $\chi_R$ whose VEV around several TeV triggers the breaking of $SU(2)_R\times U(1)_{B-L}$ down to the SM hypercharge group $U(1)_Y$, plus a bidoublet $\Sigma$ whose 
VEV triggers the $SU(2)_L\times U(1)_{Y}$ breaking at the weak scale (see~\cite{Dobrescu:2015jvn} for more details). The corresponding covariant derivatives are given by
\be
D^\mu \psi_{\chi}\,\equiv\,\derp^\mu \psi_{\chi} \, + \,\frac{i}{2}\,\gchi\,W^{\mu,a}_{\chi}\,\tau^a_\chi\,\psi_{\chi} + i\,g'\,B^\mu\,Y_{B-L}\,\psi_{\chi}\,, 
\label{Fermion-Covariant-derivatives}
\ee

\nt where $\tau^a_\chi$ and $Y_{B-L}$ correspond to the $SU(2)_\chi$ and $U(1)_{B-L}$ generators,  with $\chi\equiv L,\,R$, and the fermion field $\psi$ standing for $\psi\equiv q,\,l$. Other fermion arrangements, dictated either by leptophobic, hadrophobic, fermionphobic~\cite{FP0,LH4,FP}, ununified~\cite{UnunifiedSM} or non-universal~\cite{NU} are also possible and are beyond the scope of this work.

Operators mixing the LH and RH-covariant are also constructable in this approach via the proper insertions of the Goldstone matrices $\UHL$ and $\UHR$, more specifically, through the following definitions~\cite{Yepes:2015zoa}
\be
\widetilde{\VL}^\mu_\chi \equiv \UH^\dagger_\chi\,\VLchimuu\,\UH_\chi,\, 
\qquad 
\widetilde{\TL}_\chi \equiv \UH^\dagger_\chi\,\TLchi\,\UH_\chi\,,
\label{Vtilde-Ttilde}
\ee
\be
\WWchiut \equiv \UH^\dagger_\chi\,\WWchiu\,\UH_\chi\,,
\label{Wtilde}
\ee
\nt where $\WWchiu\equiv W^{\mu\nu,a}_\chi\tau^a/2$. Non--zero NP departures with respect to those described in $\LL_0\,+\,\LL_{0,R}\,+\,\LL_{0,LR}$ will be parametrized through the remaining last two pieces in~\eqref{Lchiral}, i.e. $\Delta\LL_{\text{CP}}$ and $\Delta\LL_{\text{CP},LR}$. The former contains LH and RH covariant objects up to the $p^4$--order as
\be
\Delta \LL_{\text{CP}}=\Delta \LL_{\text{CP},L}+\Delta \LL_{\text{CP},R}\,.
\label{DeltaL-CP-even}
\ee

\nt The latter can be further written down as 
\beq
\hspace*{-1.5mm}
\Delta \LL_{\text{CP},L}=\alpha_B\cP_B(h)+\sum_{i=\{W,C,T\}}\alpha_i\cP_{i,L}(h)+\sum_{i=1}^{26} \alpha_i\cP_{i,L}(h)
\label{DeltaL-CP-even-L}
\eeq
\be
\Delta\LL_{\text{CP},R}= \sum_{i=\{W,C,T\}}\beta_i\cP_{i,R}(h)\,\, \,+\,\,\,\sum_{i=1}^{26} \beta_i\cP_{i,R}(h)\,.
\label{DeltaL-CP-even-R}
\ee

\nt The model--dependent constant coefficients $\alpha_i$ and $\beta_i$ are denoting correspondingly the weighting coefficients for the LH and RH operators, whilst the first two terms of $\Delta \LL_{\text{CP},L}$ in~\eqref{DeltaL-CP-even-L} and the first term in~\eqref{DeltaL-CP-even-R} can be jointly written as
\beq
\begin{aligned}
\cP_B(h)\,\,  &= -\frac{g'^2}{4}\,B_{\mu\nu}\,B^{\mu\nu}\,\cF_B(h)\,, \\ 
\cP_{W,\,\chi}(h)\,\,  &= -\frac{\gchi^2}{4}\, W_{\mu\nu,\,\chi}^a\,W^{\mu\nu,\,a}_\chi\,\cF_{W,\,\chi}(h)\,,  \\  
\cP_{C,\,\chi}(h)\,\,  &= - \frac{\fchi^2}{4}\Tr\Big(\VLchimuu\,\VLchimud\Big) \,\cF_{C,\,\chi}(h)\,, \\ 
\cP_{T,\,\chi}(h)\,\,  &= \frac{\fchi^2}{4}\, \Big(\Tr\Big(\TLchi\,\VLchimuu\Big)\Big)^2\,\cF_{T,\,\chi}(h)\,,  
\label{GT}
\end{aligned}
\eeq    

\nt with suffix $\chi$ labelling again as $\chi=L,R$, and the generic $\cF_i(h)$--function of the scalar singlet $h$ is introduced for all the operators following definition~\eqref{F}. No gluonic operator has been included in $\Delta \LL_{\text{CP},L}$. The contribution $\Delta \LL_{\text{CP},L}$ has already been provided in~\cite{Alonso:2012px,Brivio:2013pma} in the context of purely EW chiral effective theories coupled to a light Higgs, whereas part of $\Delta \LL_{\text{CP},L}$ and $\Delta \LL_{\text{CP},R}$ were partially analysed for the left--right symmetric frameworks in~\cite{Zhang:2007xy,Wang:2008nk}, and finally completed in~\cite{Yepes:2015zoa}. 

Finally, $\Delta\LL_{\text{CP},LR}$ parametrizes any possible mixing interacting term between the $SU(2)_L$ and $SU(2)_R$--covariant objects up to the $p^4$--order in the Lagrangian expansion, permitted by the underlying left--right symmetry, and encoded through
\be
\hspace*{-1mm}
\Delta\LL_{\text{CP},LR}=\sum_{i=\{W,C,T\}}\gamma_i\cP_{i,LR}(h)\,+\sum_{i=2,\,i\neq 4}^{26} \gamma_{i(j)}\,\cP_{i(j),LR}(h)
\label{DeltaL-CP-even-LR}
\ee

\nt where the index $j$ spans over all the possible operators that can be built up from the set of 26 operators $\cP_{i,\chi}(h)$ in~\eqref{DeltaL-CP-even-L}--\eqref{DeltaL-CP-even-R}, and here labelled as $\cP_{i(j),LR}(h)$ together with their corresponding coefficients $\gamma_{i(j)}$. The first term in $\Delta\LL_{\text{CP},LR}$ encodes the non-linear mixing operators
\be
\hspace*{-1.5mm}
\begin{aligned}
\cP_{W,\,LR}(h)\,\,  &= -\frac{1}{2}\,\gL\,\gR\,\Tr\left(\WWLut\,\WWRdt\right)\,\cF_{W,\,LR}(h)\,,  \\ \\
\cP_{C,\,LR}(h)\,\,  &=  \frac{1}{2}\,\fL\,\fR\,\Tr\Big(\VLLmuut\VLRmudt\Big)\,\cF_{C,\,LR}(h)\,, \\ \\
\cP_{T,\,LR}(h)\,\,  &= \frac{1}{2}\,\fL\,\fR\,\Tr\Big(\TLLt\,\VLLmuut\Big)\,\Tr\Big(\TLRt\,\VLRmudt\Big)\cF_{T,\,LR}(h)
\label{WT-LR}
\end{aligned}
\ee

\nt The complete set of operators $\cP_{i(j),LR}(h)$ in the second term of $\Delta\LL_{\text{CP},LR}$ have been fully and listed in~\cite{Yepes:2015zoa}. The corresponding CP--violating counterparts of $\Delta\LL_{\text{CP}}$ and $\Delta\LL_{\text{CP},LR}$ have been completely listed and studied in~\cite{Yepes:2015qwa}. Notice that in the unitary gauge, non-zero mass mixing terms among the LH and RH gauge fields are triggered by the operator $\cP_{C,\,LR}(h)$, leading to diagonalize the gauge sector in order to obtain the required physical gauge masses.

\vspace*{0.7cm}
\subsection{Charged and neutral gauge masses}
\label{Charged-neutral-masses}

\nt The gauge basis is defined by
\be
\widehat{\cW}_{\mu }^\pm\equiv 
\left(
\begin{array}{c}
 W_{\mu,L}^\pm \\ [3mm]
 W_{\mu,R}^\pm \\
\end{array}
\right)\,, \qquad\qquad 
\widehat{\cN}_{\mu }\equiv \left(
\begin{array}{c}
 W_{\mu ,L}^3 \\ [3mm]
 W_{\mu ,R}^3 \\[3mm]
 B_{\mu } \\
\end{array}
\right)
\label{Gauge-basis}
\ee

\nt where the charged fields $W_{\mu,\chi}^\pm$ are introduced as usual
\be
W_{\mu,\chi}^\pm\equiv \frac{W_{\mu,\chi}^1\,\mp\,i\,W_{\mu,\chi}^2}{\sqrt{2}}\,,\qquad\qquad \chi=L,\,R\,.
\label{Charged-fields}
\ee

\nt The mass eigenstate basis is defined as
\be
\cW_{\mu }^\pm\equiv 
\left(
\begin{array}{c}
W_{\mu}^\pm \\[3mm]
W^{'\pm}_{\mu} 
\end{array}
\right)\,, \qquad\qquad 
\cN_{\mu }\equiv \left(
\begin{array}{c}
A_{\mu} \\[3mm]
Z_{\mu} \\[3mm]
Z'_{\mu} \\
\end{array}
\right)\,,
\label{Mass-basis}
\ee

\nt and it can be linked to the gauge basis through the following field transformations 
\be
\widehat{\cW}_{\mu }^\pm\equiv 
\cR_{\cW}\,\cW_{\mu }^\pm\,, \qquad\qquad \widehat{\cN}_{\mu }\equiv \cR_{\cN}\,\cN_{\mu }\,.
\label{Gauge-field-transformations}
\ee

\nt The mass matrices for the charged and neutral sector in the gauge 
basis are
\be
\cM_{\cW}=
 \frac{\gL^2\,\fL^2}{4}\,\left(
\begin{array}{cc}
1+\alpha_C & -\frac{\gamma_C}{\sqrt{\lambda }}  \\ \\
- \frac{\gamma_C}{\sqrt{\lambda }} & \frac{\left(1+\beta_C\right)}{\lambda } \\
\end{array}\right)\,,\quad \lambda\equiv\frac{\gL^2}{\gR^2}\,\epsilon^2,\,\quad\epsilon\equiv\frac{\fL}{\fR}\,.
\label{Charged-gauge-matrix}
\ee

\begin{widetext}
\be
\small{
\begin{aligned}
&\cM_{\cN}=\frac{g_L^2 \fL^2}{4}\left(
\begin{array}{ccc}
1+\alpha & -\frac{\gamma}{\sqrt{\lambda}} & -\frac{g'}{\gL}
\left(1+\alpha-\frac{\fR}{\fL}\gamma\right) \\ \\
 -\frac{\gamma}{\sqrt{\lambda }} & \frac{1+\beta}{\lambda } & \frac{g'}{\gL\sqrt{\lambda }}\left(\gamma-\frac{\fR}{\fL}\left(1+\beta\right)\right) \\ \\
-\frac{g'}{\gL}
\left(1+\alpha-\frac{\fR}{\fL}\gamma\right) & \frac{g'}{\gL\sqrt{\lambda }}\left(\gamma-\frac{\fR}{\fL} \left(1+\beta\right)\right) & \frac{g'^2 }{\gL^2 }
\left(1+\alpha-\frac{2\fR}{\fL}\gamma +\frac{\fR^2}{\fL^2}\left(1+\beta\right)\right) \\ \\
\end{array}
\right)
\label{Neutral-gauge-matrix}
\end{aligned}
}
\ee

\nt with the definitions
\be
\alpha \equiv \alpha_C-2\,\alpha_T\,,\qquad\qquad
\beta \equiv \beta_C-2\,\beta_T\,,\qquad\qquad
\gamma \equiv \gamma_C+2\,\gamma_T\,.
\label{alpha-beta-gamma}
\ee

\end{widetext}

\nt The rotation matrix for the charged sector can be written down as
\be	
\cR_{\cW}=
\left(
\begin{array}{cc}
 c_\zeta & -s_\zeta \\ \\
 s_\zeta & c_\zeta \\
\end{array}
\right)\,,
\qquad c_\zeta\equiv \cos\zeta\,,
\quad  s_\zeta\equiv \sin\zeta\,. 
\label{Charged-rotation}
\ee

\nt For the neutral sector the rotation is dictated by the Euler-angles parametrization in terms of three angles: the Weinberg mixing angle $\theta_W$, and the analogous mixing angle $\theta_R$ for the $SU(2)_R\times U(1)_{B-L}$ subgroup, both defined as
\be
\cos\theta_W\equiv c_W = \frac{\gL}{\sqrt{\gL^2 + g^2_Y}},\,\quad \sin\theta_W\equiv s_W = \frac{g_Y}{\sqrt{\gL^2 + g^2_Y}}
\label{SM-mixing-angle}
\ee
\be
\cos\theta_R\equiv c_R = \frac{\gR}{\sqrt{\gR^2 + g'^2}},\,\quad \sin\theta_R\equiv s_R = \frac{g'}{\sqrt{\gR^2 + g'^2}}\,.
\label{Right-B-L-mixing-angle}
\ee

\nt The third angle $\phi$ can be linked to the latter two up to $\cO(\epsilon^2\,\gamma^2)$--contributions through
\be
\tan\phi \,\,\simeq\,\,\epsilon\,\frac{g_L}{g_R}\,\frac{c_R}{c_W}\left(\epsilon\,s^2_R-\gamma\right)\,.
\label{Third-mixing-angle}
\ee

\nt The rotation matrix for the neutral sector becomes parametrized then as
\be
\hspace*{-3mm}
\small{
\cR_{\cN}\simeq
\left(
\begin{array}{ccc}
 s_W & c_W & \epsilon\,c_R\,\frac{g_L}{g_R}\left(\gamma -\epsilon\,s_R^2\right) \\ \\
 c_W\,s_R & -s_R\, s_W -\frac{g_L}{g_R}\frac{c_R^2}{c_W}\,\epsilon\,\gamma & c_R \left(1-\frac{g_L}{ g_R}\frac{s_R s_W}{c_W}\,\epsilon\,\gamma\right) \\ \\
c_R\,c_W & c_R \left(-s_W + \frac{g_L}{g_R}\frac{s_R}{c_W}\,\epsilon\,\gamma  \right) & -s_R - \frac{g_L}{g_R}\frac{s_W c_R^2}{c_W}\,\epsilon\,\gamma \\
\end{array}
\right)\,.
}
\label{Neutral-rotation}
\ee

\nt with the coefficient $\gamma$ encoding the contributions induced by the left--right custodial conserving and custodial breaking operators $\cP_{C,LR}(h)$ and $\cP_{T,LR}(h)$ respectively (defined in~\eqref{alpha-beta-gamma}). Such contributions are suppressed by the scale ratio $\epsilon$. In the limit $\fL\ll\fR$, the charged gauge masses are
\be
\begin{aligned}
M^2_W &\simeq \frac{1}{4}\,\gL^2\,\fL^2\,\Big(1 - \lambda\,\gamma^2_C\Big),\,\quad M^2_{W^\prime} \simeq \frac{1}{4}\,\gR^2\,\fR^2\,\Big(1 + \lambda\,\gamma^2_C\Big)\,.
\label{Charged-masses-expanded}
\end{aligned}
\ee

\nt where the masses have been expanded up to $M^2_W/M^2_{W'}$-terms. The mixing angle $\zeta$ for the charged sector turns out to be depending on the masses ratio $M^2_W/M^2_{W'}$ through the parameter $\lambda$ and the mixing coefficient $\gamma_C$ in~\eqref{Charged-gauge-matrix} as 
\be
\tan\zeta= -\frac{\sqrt{\lambda }}{1-\lambda}\, \gamma_C\,,\qquad 
\lambda\equiv\frac{\gL^2\,\fL^2}{\gR^2\,\fR^2}\simeq \frac{M^2_W}{M^2_{W'}}\,.
\label{Charged-Mixing-angle}
\ee

\nt The neutral gauge masses are
\be
M^2_Z\,\simeq\,\frac{M^2_W}{c^2_W}\,,\qquad
M^2_{Z^\prime}\,\simeq\,\frac{M^2_{W'}}{c^2_R}\left(1-2\,s^2_R\,\epsilon\,\gamma\right)\,
\label{Neutral-masses-expanded}
\ee

\nt with the coefficient $\gamma$ introduced in~\eqref{alpha-beta-gamma}. The well measured $M_Z$--mass strongly constrains additional contributions from the operators $\cP_{C,\,L}(h)$ and $\cP_{T,\,L}(h)$ in~\eqref{Neutral-masses-expanded}. Similarly, the $M_W$--mass bounds tightly constrains the contribution of $\cP_{C,\,R}(h)$ in~\eqref{Charged-masses-expanded}. 

As it can be noticed from~\eqref{Neutral-masses-expanded}, the $Z'$-mass turns out to be larger with respect to the $W'$-mass, i.e  $M_{Z^\prime}>M_{W^\prime}$. In addition, a mass range for the neutral gauge field $Z'$ can be predicted in terms of the $W'$--mass and the gauge couplings $\gR$ and $g_Y$, via the mixing angle $\theta_R$ in~\eqref{Right-B-L-mixing-angle} and the link among the $SU(2)_L$, $U(1)_{B-L}$ and the SM hypercharge gauge couplings as
\be
\frac{1}{g_R^2} + \frac{1}{g'^2}= \frac{1}{g_Y^2}\,.
\label{B-L-gauge-coupling}
\ee 

\nt The observed excess at the ATLAS and CMS Collaborations around invariant mass  of 1.8--2 TeV can be interpreted to be induced by a
$W'$--contribution. The coupling $\gR$ will determine the strength of the couplings among the $W'$ and fermions fields, and therefore it will control as well the production rate of $W'$--resonances via the process $p\,p \to W^\prime$ analysed in the following section.


\section{$W'$--production}
\label{Diboson-excess-W'-production}

\nt By considering the charged currents from the Lagrangians $\LL_0$ and $\LL_{0,R}$ in~\eqref{LLO} and~\eqref{LLO-Right} respectively, we have
\be
\LL_{ud W'}= -\frac{1}{\sqrt{2}}\,{\bar u}\,\gamma^\mu\left(g_L\,\sqrt{\lambda }\,\gamma_C\,P_L - \gR\,P_R\right)d\,W'_\mu\,\,+\,\,\hc\,, 
\label{W'-charged-currents}
\ee

\begin{figure}
\begin{center}
\includegraphics[scale=0.59]{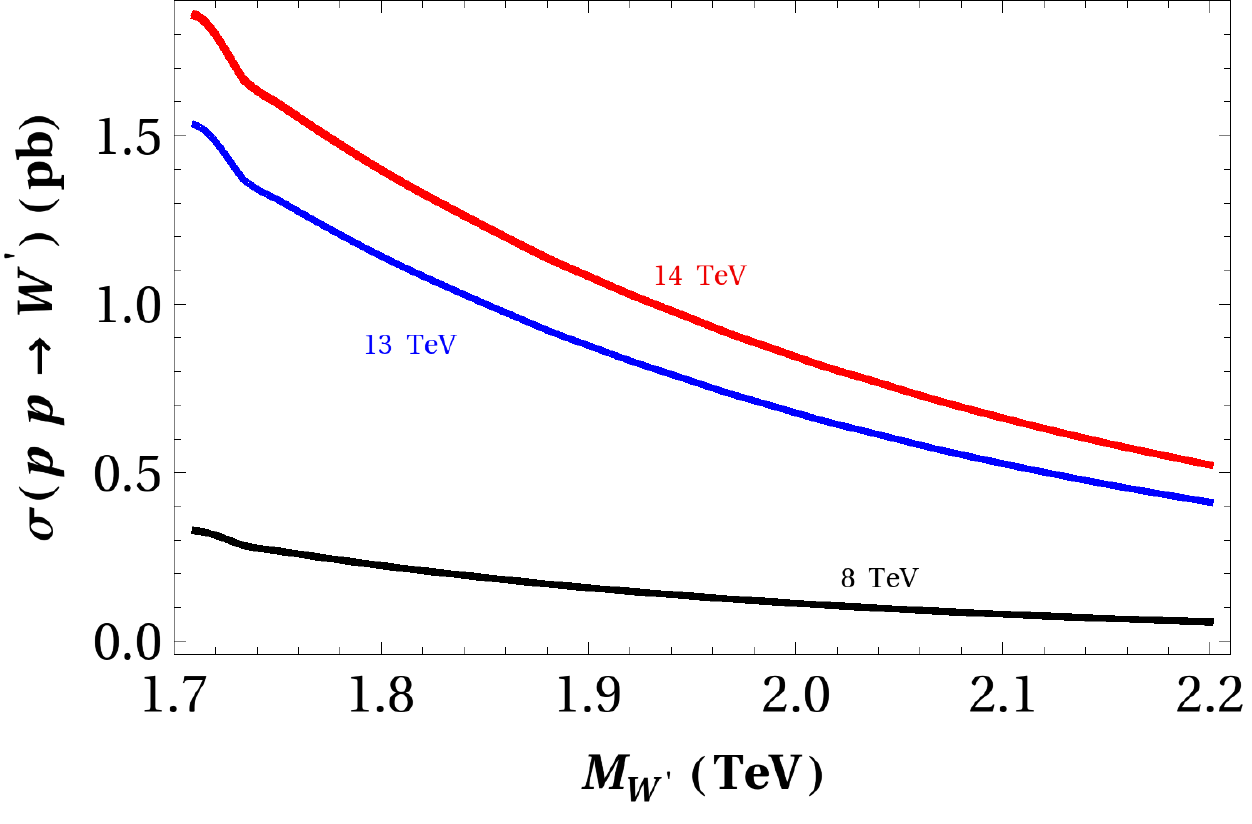}
\caption{\sf $W'$-production cross section via the process $p\,p \to W^\prime$ as a function of $M_{W'}$, for $\gR=0.5$ and at the 8-13-14 TeV LHC (black, blue and red curves respectively). Departures with respect to the vanishing $\gamma_C$-case are suppressed by $M_W/M_{W'}$ and can be entirely neglected from the production cross section.}
\label{pp-W'-production}
\end{center}
\end{figure}

\nt where a flavour diagonal couplings have been assumed and the family indices are implicit, with $P_{L(R)}\equiv \left(1\mp\gamma^5\right)/2$. The $W^\prime$-production cross section through the process $p\,p \to W^\prime$ can be computed from the Lagrangian in~\eqref{W'-charged-currents} by using MadGraph 5 and implementing the scale-dependent $K$-factors calculated in~\cite{Cao:2012ng}. They are in the ranges $K\,\in\,[1.32,\,1.37]$ at $\sqrt{s} = 8$ TeV and $K\,\in\,[1.23,\,1.25]$ at 13-14 TeV. Fig.~\ref{pp-W'-production} shows the $W^\prime$-production cross section for $\gR=0.5$ at the center-of-mass (c.o.m) energies 8-13-14 TeV LHC (black, blue and red curves respectively). The coefficient $\gamma_C$ is running as $\gamma_C=-1.0,\,0,\,1$. In general, departures with respect to the vanishing $\gamma_C$-case are suppressed by the ratio $\sqrt{\lambda }\simeq M_W/M_{W'}$, and they can be neglected for the $W^\prime$-production. As it can be seen from Fig.~\ref{pp-W'-production}, the cross section productions are

\begin{itemize}
\item At $M_{W'}\sim\!1.8\,\text{TeV}$, around $\sim\!0.25\,\text{pb},\,1.2\,\text{pb},\,1.5\,\text{pb}$ at $\sqrt{s}\,=\,$8-13-14 TeV respectively;
\item At $M_{W'}\sim\!2\,\text{TeV}$, around $\sim\! 0.13\,\text{pb},\,0.7\,\text{pb},\,0.9\,\text{pb}$ and at the same c.o.m energies correspondingly.
\end{itemize}

\nt The coupling $g_R$ can be determined from the cross section required to produce the dijet resonance near $M_{W'}$.
The  CMS dijet excess \cite{Khachatryan:2015sja} 
at a mass in the 1.8--1.9 TeV range indicates that the $W'$ production cross section times the dijet branching fraction 
is in the 100--200 fb range (this is consistent with the ATLAS dijet result \cite{Aad:2014aqa}, which shows a smaller excess at 1.9 TeV). 
This was assumed  in Refs.~\cite{Dobrescu:2015qna,Brehmer:2015cia}
to be the range for $\sigma ( pp \to W' \! \to jj)$, where $j$ is a hadronic jet associated with quarks or antiquarks other than the top. By comparing the $W^\prime$ production cross section to the CMS dijet excess, the coupling $\gR$ was determined in the range $\gR \approx 0.45-0.6$~\cite{Dobrescu:2015qna}. A similar range is obtained by computing the dijet decay channel of a $W'$ in our scenario, and it will be assumed henceforth. Such range, together with a $W^\prime$--boson mass nearby 1.8--2 TeV, can be translated via the $W^\prime$ mass formula in~\eqref{Charged-masses-expanded} into the relation 
\be
\fR\,\approx\,\frac{3.6\text{--}4\,\text{TeV}}{\gR}\,\approx\,6-8\,\text{TeV}\,.
\label{fR-rough-value}
\ee

\nt The $W'$-production via the decay modes $p\,p \to W' \to WZ$ and $p\,p \to W' \to W h$, together with the observed excesses in the $WZ$ and $Wh$--final states at ATLAS and CMS, allow us to infer ranges for the strength of the associates operators contributing to those channels. The latter can be described by the effective Lagrangians
\begin{widetext}
\be
\LL_{W W' Z} = \\ i\left(\mathit{g}_{\text{\textit{$W W' Z$}}}^{(1)}
\,W^\dag_{\mu\nu}\,W^{'\nu}\,Z^\mu \,\, + \,\, \mathit{g}_{\text{\textit{$W W' Z$}}}^{(2)}
\,W^{'\dag}_{\mu\nu}\,W^\nu\,Z^\mu  \,\, + \,\, \mathit{g}_{\text{\textit{$W W' Z$}}}^{(3)}\,Z_{\mu\nu}\,W^{\mu\dag}\,W^{'\nu} \,\, + \,\,\hc\right)\,,
\label{Cubic-gauge-left-right-interactions}
\ee

\be
\LL_{h W W'} =  
- \frac{1}{M_W}\,g_{h W W'}^{(1)}\,\left(W^\dag_{\mu\nu}\,W^{'\mu\nu}\,h \,\,+\,\, \hc\right)\,\,+\,\, 
g_{ h W W'}^{(2)}\,M_W\,\left(W_{\mu}^\dag\,W^{'\mu}\,h
\,\,+\,\, \hc\right)\,,
\label{Cubic-gauge-h-left-right-interactions}
\ee
\end{widetext}

\begin{table}[htpb!]
\centering
\renewcommand{\arraystretch}{1.5}
\begin{tabular}{||c||c||} 
\hline\hline 
\multicolumn{2}{||c||}{$\bf W'\to WZ$} \\
\hline\hline
\\[-5mm]
$\mathit{g}_{\text{\textit{$W W' Z$}}}^{(1)} $    & $ \frac{e}{4\,c_W^2}  \left(\frac{e^2}{s_R}\,\gamma _W + \frac{2\,c_W}{s_W}\frac{M_W}{M_{W'}}\, \gamma _C\right)$ 
\\[3mm]
$ \mathit{g}_{\text{\textit{$W W' Z$}}}^{(2)} $    & $ -\frac{e}{4\,s_W^2} \left(\frac{e^2}{s_R}\,\gamma _W - \frac{2\,s_W}{c_W}\frac{M_W}{M_{W'}}\,\gamma _C\right) $ 
\\[3mm]
$ \mathit{g}_{\text{\textit{$W W' Z$}}}^{(3)} $   & 
$ -\frac{e}{c_W \,s_W}\frac{M_W}{ M_{W'}}\,\gamma _C $\\[1mm]
\hline \hline
\multicolumn{2}{||c||}{$\bf W'\to W h$}\\
\hline\hline
\\[-5mm]
$\mathit{g}_{\text{\textit{$h W W'$}}}^{(1)}$
 & 
$ \frac{e^3}{4\,c_W\,s_R\,s_W^2}\, \widetilde{\gamma}_W $ 
\\[3mm]
$ \mathit{g}_{\text{\textit{$h W W'$}}}^{(2)} $ & $ -\frac{e}{s_W }\frac{M_W}{M_{W'}}\left[\gamma _C + \frac{M_{W'}^2}{M_W^2}\left(\widetilde{\gamma}_C -\gamma _C \right)\right] $
\\[2mm]
\hline \hline
\end{tabular}
\caption{\sf Effective couplings encoded by the Lagrangians $\LL_{W W' Z}$ and $\LL_{h W W'}$ in~\eqref{Cubic-gauge-left-right-interactions} and~\eqref{Cubic-gauge-h-left-right-interactions} respectively. The relations $g_L=\frac{e}{s_W},\,g_R=\frac{e}{c_W\,s_R},\,g'= \frac{e}{c_R\,c_W}$ have been implemented through all the couplings, with $e$ the electromagnetic coupling constant. The coefficient $\widetilde{\gamma}_i$ stands for $\widetilde{\gamma}_i\,\equiv\,a_i\,\gamma_i$, with $i=C,\,W$ and $a_i$ coming from the $\cF(h)$--definition in~\eqref{F}.} 
\label{Cubic-gauge-left-right-couplings}
\end{table}

\nt with $V_{\mu\nu}\equiv \partial_\mu V_{\nu}-\partial_\nu V_{\mu}$, for $V \equiv W,\,W',\,Z$. The corresponding couplings are collected in Table~\ref{Cubic-gauge-left-right-couplings}. Only the LO Lagrangian $\LL_0\,+\,\LL_{0,R}$ in~\eqref{LLO}-\eqref{LLO-Right} and the operators set in~\eqref{GT} and~\eqref{WT-LR} have been kept for simplicity. Additional contributions from the operators $\cP_{i,\,L}(h)$ and $\cP_{i,\,R}(h)$ (3rd and 2nd terms in Eq.~\eqref{DeltaL-CP-even-L}-\eqref{DeltaL-CP-even-R}), and the operators $\cP_{i(j),LR}(h)$ (2nd term in~Eq.\eqref{DeltaL-CP-even-LR}) would lead to a larger parameter space and it is beyond the scope of this work. Many of those operators are also irrelevant at low energies as their contribution become negligible once the RH gauge filed content is integrated out from the physical spectrum~\cite{Shu:2015cxm}. We will keep henceforth the Lagrangians in~\eqref{LLO}-\eqref{LLO-Right} and the operators set in~\eqref{GT} and~\eqref{WT-LR} for the analysis below.

\subsection{$WZ$ and $W h$ excesses}
\label{WZ-Wh-excess}

\nt For a charge resonance around the TeV scale, the ratios $M^2_Z/M^2_{W'}$ and $M^2_H/M^2_{W'}$ turns out to be negligible and therefore the decay width for the processes $W' \to WZ$ and $W' \to W h$ become written as
\be
\Gamma\left(W' \to WZ\right)=\frac{c_W^2}{192\,\pi}\frac{M_{W'}^5}{M_W^4}\,\left(\mathit{g}_{\text{\textit{$W W' Z$}}}^{(2)}\right)^2\,,
\label{W'-WZ-width-limiting-case}
\ee
\be
\Gamma\left(W' \to W h\right)=\frac{g_{\text{\textit{$h$}}\text{\textit{$ $}}WW'}^{\text{(1)}}}{48\,\pi}\left(g_{\text{\textit{$h$}}\text{\textit{$ $}}WW'}^{\text{(1)}} + \,g_{\text{\textit{$h$}}\text{\textit{$ $}}WW'}^{\text{(2)}}\,\frac{M_W^2}{M_{W'}^2}\right)\frac{M^5_{W'}}{M_W^4}\,.
\label{W'-Wh-width-limiting-case}
\ee

\nt The cross sections for the processes $p\,p \to W' \to WZ$ and $p\,p \to W' \to Wh$ can be computed in terms of the corresponding one for the decay $p\,p \to W' \to jj$ as
\be
\frac{\sigma_{WZ}(W')}{\sigma_{jj}(W')} = \frac{\Gamma (W'\to WZ)}{\Gamma (W'\to jj)},\quad 
\frac{\sigma_{Wh}(W')}{\sigma_{jj}(W')} = \frac{\Gamma (W'\to Wh)}{\Gamma (W'\to jj)}   
\label{Cross-sections-ratio}
\ee

\nt with $\sigma_{XX}(W')  \equiv \sigma ( pp \to W' \! \to XX)$. Neglecting the $M_W/M_{W'}$--corrections induced by the operators $\cP_{C,LR}(h)$ and $\cP_{T,LR}(h)$ (see Eq.~\eqref{W'-charged-currents}), the width for the decay $W'\to jj$ can be related to the process $W'\to t\bar b$ through the Lagrangian in~\eqref{W'-charged-currents} as 
\be
\begin{aligned}
\Gamma (W'\to jj) &\simeq 2\,\Gamma(W'\to t \bar b) \sim  \frac{\gR^2}{8\,\pi}\,M_{W'}\,.
\label{Width-cross-section-jets}
\end{aligned}
\ee

\begin{table}[htpb!]
\renewcommand{\arraystretch}{0.7}
\small{
\begin{tabular}{cc||c||c}
\hline\hline
\\[-2.2mm]
\multicolumn{2}{c||}{\bf Coeff.}  & $\bf 100\,\text{fb}$  & $\bf 200\,\text{fb}$ 
\\[0.3mm]
\hline\hline
\\[-2.5mm]
\multicolumn{1}{c||}{$\gamma_C$}  & 
\multicolumn{1}{c||}{$\begin{array}{l}  
<0\\
\\[-1mm]
>0
\end{array}$} &
$\begin{array}{l}  
[-0.11,\,-0.06]\\
\\[-1mm]
\hspace*{3mm}[0.06,\,0.11]
\end{array}$
&
$\begin{array}{l}  
[-0.07,\,-0.04]\\
\\[-1mm]
\hspace*{3mm}[0.04,\,0.07]
\end{array}$
\\[4mm]
\hline
\\[-2.5mm]
\multicolumn{1}{c||}{$\gamma_W$}  & 
\multicolumn{1}{c||}{$\begin{array}{l}  
<0\\
\\[-1mm]
>0
\end{array}$} &
$\begin{array}{l}  
[-0.026,\,-0.018]\\
\\[-1mm]
\hspace*{3mm}[0.018,\,0.026]
\end{array}$
& 
$\begin{array}{l}  
[-0.018,\,-0.013]\\
\\[-1mm]
\hspace*{3mm}[0.013,\,0.018]
\end{array}$
\\[3.5mm]
\hline\hline
\end{tabular}
}
\caption{\sf Allowed negative and positive ranges for the coefficients $\gamma_C$ and  $\gamma_W$ (upper and lower rows) and for the values $\sigma_{jj}(W') \sim 100\!-\!200\,\text{fb}$~\cite{Aad:2014aqa} (3rd \& 4th columns). The values $\sigma_{WZ}(W')\sim 3\!-\!10\,\text{fb}$~\cite{Aad:2015owa}, the equivalence relation $\sigma_{Wh}(W') \approx \sigma_{WZ}(W')$ and the coefficients  $a_{C,LR}=a_{W,LR}=1/2$, were implemented for the $W' \to WZ$ and $W' \to Wh$--decay widths in~\eqref{W'-WZ-width-limiting-case}-\eqref{W'-Wh-width-limiting-case} with the relations in~\eqref{Cross-sections-ratio}-\eqref{Width-cross-section-jets}.
}
\label{CLR-CWLR-bounds-W'WZ-W'Wh}
\end{table}

\begin{figure}[htpb!]
\begin{center}
\includegraphics[scale=0.59]{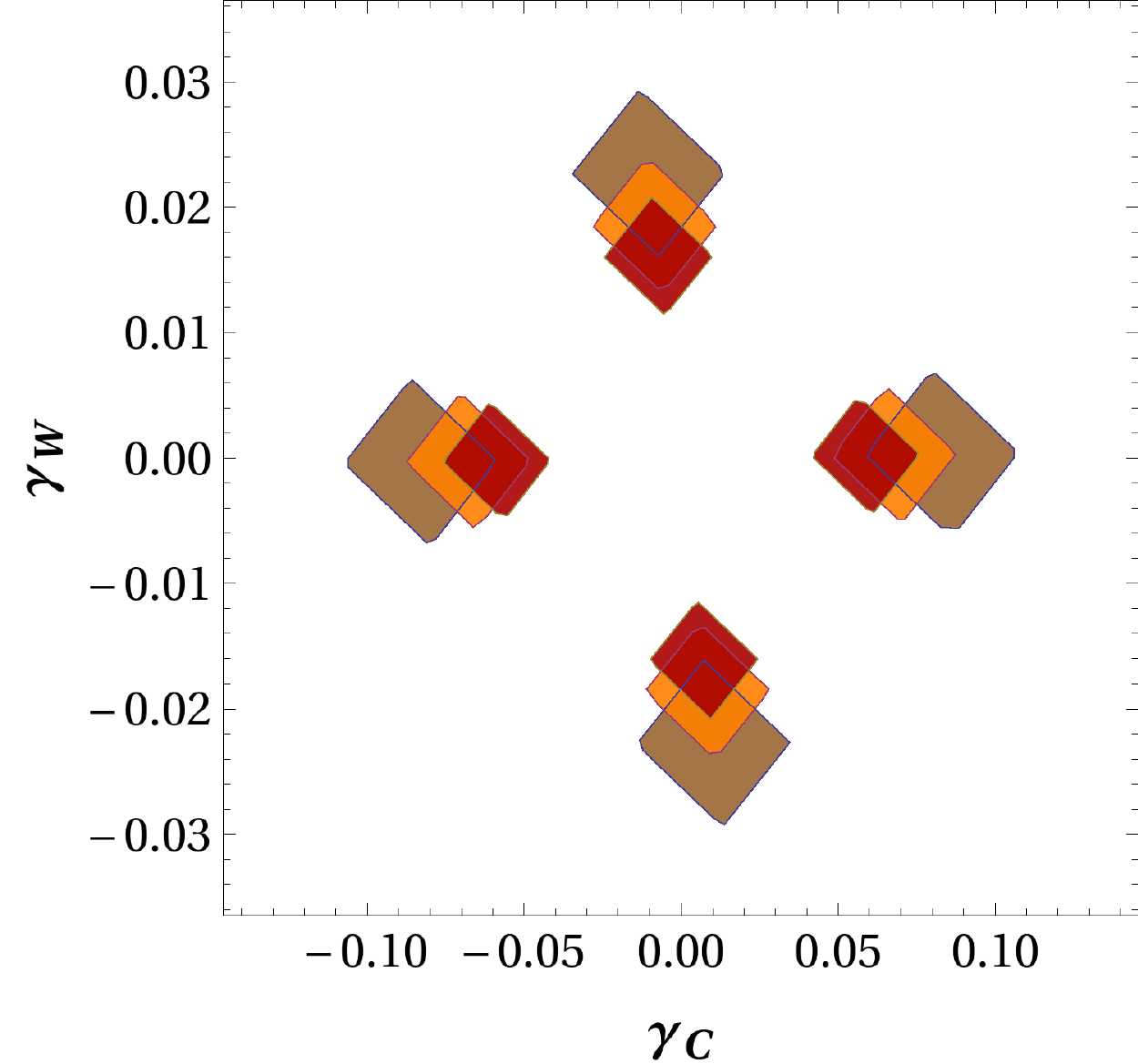}
\caption{\sf Allowed parameter space $\left(\gamma_C,\,\gamma_W\right)$ by combining the $W' \to WZ$ and $W' \to Wh$--decay widths in~\eqref{W'-WZ-width-limiting-case}-\eqref{W'-Wh-width-limiting-case} together with the relations in~\eqref{Cross-sections-ratio}-\eqref{Width-cross-section-jets}. The cross section values $\sigma_{Wh}(W') \approx \sigma_{WZ}(W') \sim 3\!-\!10\,\text{fb}$ and  $\sigma_{jj}(W') \sim 100\!-150-\!200\,\text{fb}$ (brown, orange and red charts respectively) have been implemented and assuming the Higgs coefficient values $a_{C,LR}=a_{W,LR}=1/2$.}
\label{Coefficients-parameter-space}
\end{center}
\end{figure}

\nt The Goldstone equivalence theorem requires $\Gamma (W'\! \to W h) \simeq \Gamma (W'\!\to W Z)$ up to kinematic factors. In this case the $pp \to W' \!\to Wh$ cross section satisfies $\sigma_{Wh}(W') \approx \sigma_{WZ}(W')$. Implementing in addition the results in~\eqref{W'-WZ-width-limiting-case}-\eqref{Cross-sections-ratio}, and requiring the cross section values $\sigma_{WZ}(W') \sim 3\!-\!10\,\text{fb}$ implied by the ATLAS search for $pp \! \to W' \! \to WZ \! \to JJ$~\cite{Aad:2015owa} and $\sigma_{jj}(W') \sim 100\!-\!200\,\text{fb}$~\cite{Aad:2014aqa}, we obtain the ranges for the coefficients $\gamma_C$ ($\gamma_W=0$) and $\gamma_W$ ($\gamma_C=0$) in Table~\ref{CLR-CWLR-bounds-W'WZ-W'Wh} and assuming the Higgs coefficient values $a_{C,LR}=a_{W,LR}=1/2$. Letting the coefficients $\left(\gamma_C,\,\gamma_W\right)$ to vary simultaneously, we obtain the allowed parameter space in Fig~\ref{Coefficients-parameter-space}. The ranges are basically of the same order of magnitude suggested by the ranges $-0.02<\gamma_C<0.02$ and $-0.016<\gamma_W<0.018$ obtained from the stringent EW constrains on the $Z$-gauge masses and the $S$ and $T$ parameter bounds in~\cite{Shu:2015cxm} respectively. 

It is worth to point out the dependence of the ranges in Table~\ref{CLR-CWLR-bounds-W'WZ-W'Wh} and the parameter space in Fig~\ref{Coefficients-parameter-space} on the Higgs coefficients $a_{C,LR}=a_{W,LR}=1/2$ entering in the $hWW'$--couplings through the light Higgs function in~\eqref{F}. Larger values $a_{C,LR}=a_{W,LR}\sim 1$ will reduce (enhance) the allowed positive (negative) ranges of $\gamma_W$ by one order of magnitude with respect to those in Table~\ref{CLR-CWLR-bounds-W'WZ-W'Wh} in the range $\sigma_{jj}(W') \sim 150\!-\!200\,\text{fb}$, whereas part of the ranges of $\gamma_C$ will be slightly modified and some other can reach smaller values close to zero for small values of $\gamma_W$. The limiting case $a_{C,LR}=a_{W,LR}\sim 0$ enhances the $\gamma_W$--ranges instead, but keeping the same order of magnitude of the ranges in Table~\ref{CLR-CWLR-bounds-W'WZ-W'Wh} though.

\section{$Z'$--predictions}
\label{Z'--predictions}

\nt A mass prediction for the neutral gauge field $Z'$ can be inferred from the relation~\eqref{Neutral-masses-expanded} in terms of the $W'$--mass and the gauge couplings $\gR$ and $g_Y$, via the mixing angle $\theta_R$ in~\eqref{Right-B-L-mixing-angle} and the relation in~\eqref{B-L-gauge-coupling}. Assuming the coupling $\gR$ in the range $\gR \approx 0.45-0.6$ as determined in~\cite{Dobrescu:2015qna} and $g_Y\sim 0.36$, it is possible to predict the mass range
\be
2.5\,\text{TeV}\,<\,M_{Z^\prime}\,<\,4\,\text{TeV}\,. 
\label{Z'-mass-prediction}
\ee

\nt The prospectives in detecting a $Z'$-signal in the futures collider experiments can be tackled through the fermionic decay channels  $Z^\prime \to \{\nu_L \bar\nu_L,\,N_R\bar N_R,\,\ell^+\ell^-,\,t\bar t,\, jj\}$, and via the gauge-scalar modes $Z^\prime \to \{W^+W^-,\, Z h\}$ as well, and will be analysed in the following section.

\subsection{$Z'$-production decay modes}
\label{Z'-decay-modes}

\nt By considering the neutral currents from Lagrangians $\LL_0$ and $\LL_{0,R}$ in~\eqref{LLO} and~\eqref{LLO-Right} respectively, it is possible to describe fermionic decay modes through
\be
\hspace*{-1mm}
\LL_{ffZ'}= \sum_{f=u,d,N,\nu,e}{\bar f}\,\gamma^\mu\,\left(g_{f_L f_L Z'}\,P_L+ g_{f_R f_R Z'}\,P_R\right)\,f\,Z'_\mu\,.
\label{Z'-neutral-currents}
\ee

\nt The couplings $g_{f_L f_L Z'}$ and $g_{f_R f_R Z'}$ are listed in Table~\ref{Z'-fermion-couplings}. The self gauge and gauge-Higgs Lagrangians accounting for the gauge--scalar modes will be described by 
\be
\hspace*{-2mm}
\begin{aligned}
&\LL_{W W Z'} =\\ \\
& i\left(\mathit{g}_{\text{\textit{$W W Z'$}}}^{(1)}
\,W^\dag_{\mu\nu}\,W^{\nu}\,Z^{\prime\mu}\,\, + \,\,\hc\right)
\,\, + \,\, i\mathit{g}_{\text{\textit{$W W Z'$}}}^{(2)}\,Z^{\prime}_{\mu\nu}\,W^{\mu\dag}\,W^{\nu},
\label{Cubic-gauge-left-right-interactions-Z'}
\end{aligned}
\ee

\be
\hspace*{-1mm}
\LL_{h Z Z'} =- \frac{1}{2M_Z}\,g_{h Z Z'}^{(1)}\,Z_{\mu\nu}\,Z^{'\mu\nu}\,h\,\,+\,\, 
\frac{g_{ h Z Z'}^{(2)}}{2}\,M_Z\,Z_{\mu}\,Z^{'\mu}\,h\,.
\label{Cubic-gauge-h-left-right-interactions-Z'}
\ee

\begin{table}[htb!]
\centering
\small{
\hspace*{-3mm}
\renewcommand{\arraystretch}{1.0}
\begin{tabular}{c@{\hspace*{2mm}}||c||c}
\hline\hline
\\[-3mm]
\bf f & $\bf g_{f_L f_L Z'}$ & $\bf g_{f_R f_R Z'}$
\\[0.5mm]
\hline\hline
\\[-3mm]   
$u$ &  $ \frac{e}{6 c_W}\left(\frac{s_R}{c_R}-\gamma\frac{M_W}{M_{Z'}}\frac{2 c_{2 W}+1}{c_W\,s_W}\right) $  &  $ \frac{e}{6 c_W} \left(\frac{s_R}{c_R}-\frac{3 c_R}{s_R}+4 \gamma\frac{  M_W}{M_{Z'}}\frac{s_W}{c_W}\right) $\\[4mm] 
$d$ &  $ \frac{e}{6 c_W}\left(\frac{s_R}{c_R} + \gamma\frac{M_W}{M_{Z'}}\frac{c_{2 W}+2}{c_W\,s_W}\right) $  &  $ \frac{e}{6 c_W} \left(\frac{c_{2 R}+2}{c_R s_R}-2\gamma\frac{M_W}{M_{Z'}}\frac{s_W}{c_W}\right) $\\[2mm]  
\hline
\\[-2.5mm]  
$N$ &  $0$  &  $-\frac{e}{2\,c_R\,c_W\,s_R}$\\[4mm]  
$\nu$ &  $ -\frac{e}{2 c_W}\left(\frac{s_R}{c_R} + \gamma\frac{M_W}{M_{Z'}}\frac{1}{c_W\,s_W}\right) $  &  $0$\\[4mm]  
$e$ &  $ -\frac{e}{2 c_W}\left(\frac{s_R}{c_R} - \gamma\,\frac{M_W}{M_{Z'}}\frac{c_{2 W}}{c_W\,s_W}\right) $  &  $ \frac{e}{2 c_W} \left(\frac{c_R}{s_R}-\frac{s_R}{c_R} -2 \gamma\frac{M_W}{M_{Z'}}\frac{s_W}{c_W}\right)$\\[2mm]  
\hline \hline
\end{tabular}
\caption{\sf $Z'$-fermion-couplings from the Lagrangian in~\eqref{Z'-neutral-currents}. The relation $Q=\frac{1}{2}\,T^{3}_L+\frac{1}{2}\,T^{3}_R + Y_Q$, with $T^{3}_{L(R)}\equiv \frac{1}{2}\,\tau^{3}_{L(R)}$, emerges naturally from the fermion--photon coupling in our scenario and it has been employed in all the listed couplings. In addition, the relations $g_L=\frac{e}{s_W},\,g_R=\frac{e}{c_W\,s_R},\,g'= \frac{e}{c_R\,c_W}$ have also been used, with $e$ the electromagnetic coupling constant. Notation $c_{2 W}\equiv
\cos \left(2\,\theta_W\right)$ and $c_{2 R}\equiv
\cos \left(2\,\theta_R\right)$ is implicit. 
}  
\label{Z'-fermion-couplings}
}
\end{table}

\begin{table}[htpb!]
\centering
\small{
\renewcommand{\arraystretch}{1.5}
\begin{tabular}{||c||c||} 
\hline \hline
\multicolumn{2}{||c||}{$\bf Z'\to W^+\,W^-$} \\
\hline\hline
\\[-5mm]
$\mathit{g}_{\text{\textit{$W W Z'$}}}^{(1)} $    & $ \frac{e}{2\,c_W}\frac{M_W}{M_{Z'}} \left(\frac{M_W}{M_{Z'}}\frac{s_R}{c_R}-\gamma\,\frac{c_W}{s_W}\right) $ 
\\[4mm]
$ \mathit{g}_{\text{\textit{$W W Z'$}}}^{(2)} $    & $ \frac{e}{c_W}\left[\gamma _W\,\frac{e^2\,c_R}{2\,s_R\,s_W^2}-\frac{M_W}{M_{Z'}}\left(\frac{M_W}{M_{Z'}}\frac{s_R}{c_R} - \gamma\,\frac{c_W}{s_W}\right)\right] $ 
\\[3mm]
\hline \hline
\multicolumn{2}{||c||}{$\bf Z'\to Z\,h$}\\
\hline\hline
\\[-5mm]
$\mathit{g}_{\text{\textit{$h Z Z'$}}}^{(1)}$
 & $ \frac{e^3\,c_R}{2\,c_W\,s_R\,s_W^2}\,\widetilde{\gamma}_W $ 
\\[3mm]
$ \mathit{g}_{\text{\textit{$h Z Z'$}}}^{(2)} $ & $ \frac{2\,e}{s_W} \left[\frac{s_R\,s_W}{c_R\,c_W} + \gamma\,\frac{M_W}{M_{Z'}}\left(\frac{c_{2R}}{c_R^2}\frac{s^2_W}{c^2_W}+1\right) -\frac{M_{Z'}}{M_W}\left(\widetilde{\gamma}_C +\gamma \right)\right] $
\\[2mm]
\hline \hline
\end{tabular}
\caption{\sf Effective couplings encoded at the Lagrangians $\LL_{W W Z'}$ and $\LL_{h Z Z'}$ in~\eqref{Cubic-gauge-left-right-interactions-Z'} and~\eqref{Cubic-gauge-h-left-right-interactions-Z'} respectively. The coefficient $\widetilde{\gamma}_i$ stands for $\widetilde{\gamma}_i\,\equiv\,a_i\,\gamma_i$, with $i=C,\,W$ and $a_i$ the coefficient introduced in the $\cF(h)$--definition of~\eqref{F}.} 
\label{Cubic-gauge-left-right-couplings-Z'}
}
\end{table}

\nt The corresponding couplings are collected in Table~\ref{Cubic-gauge-left-right-couplings-Z'}. Contributions induced by the left--right custodial conserving operator $\cP_{C,LR}(h)$ and the custodial breaking $\cP_{T,LR}(h)$ (encoded by the coefficient $\gamma$) are suppressed by the masses ratio $M_W/M_{Z'}$ for all the $Z'$--fermion couplings in Table~\ref{Z'-fermion-couplings}. Such contributions turn out to be suppressed 
by one factor of $M_W/M_{Z'}$ less with respect to the leading order terms for the pure gauge and gauge--Higgs couplings in Table~\ref{Cubic-gauge-left-right-couplings-Z'}, but for the coupling $ \mathit{g}_{\text{\textit{$h Z Z'$}}}^{(2)} $, whose last term is enhanced by $M_{Z'}/M_W$ due to the longitudinal helicity components in the decay $Z'\to Z h$. On the other hand, the contributions induced by the kinetic left--right operator $\cP_{W,LR}(h)$ are not $M_W/M_{Z'}$--suppressed (couplings $ \mathit{g}_{\text{\textit{$W W Z'$}}}^{(2)} $ and $\mathit{g}_{\text{\textit{$h Z Z'$}}}^{(1)}$).  These particular features enhance the corresponding leading order branching ratios of $Z'\to W^+W^-$ and $Z'\to Z h$ for a non--vanishing left--right operators $\{\cP_{C,LR}(h),\,\cP_{T,LR}(h),\,\cP_{W,LR}(h)\}$.

\begin{figure}[htpb!]
\begin{center}
\includegraphics[scale=0.53]{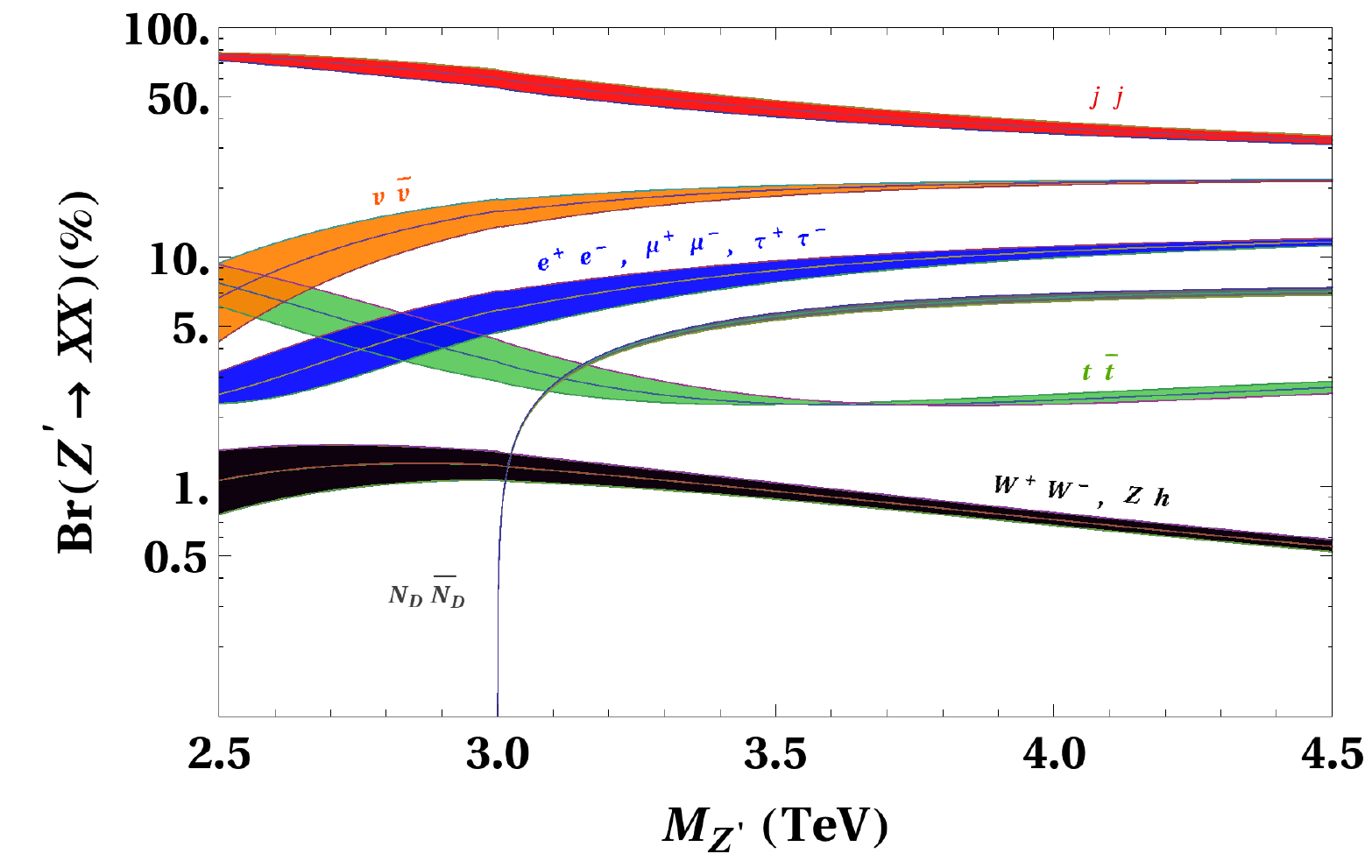}\\
\vspace*{0.5cm}
\includegraphics[scale=0.59]{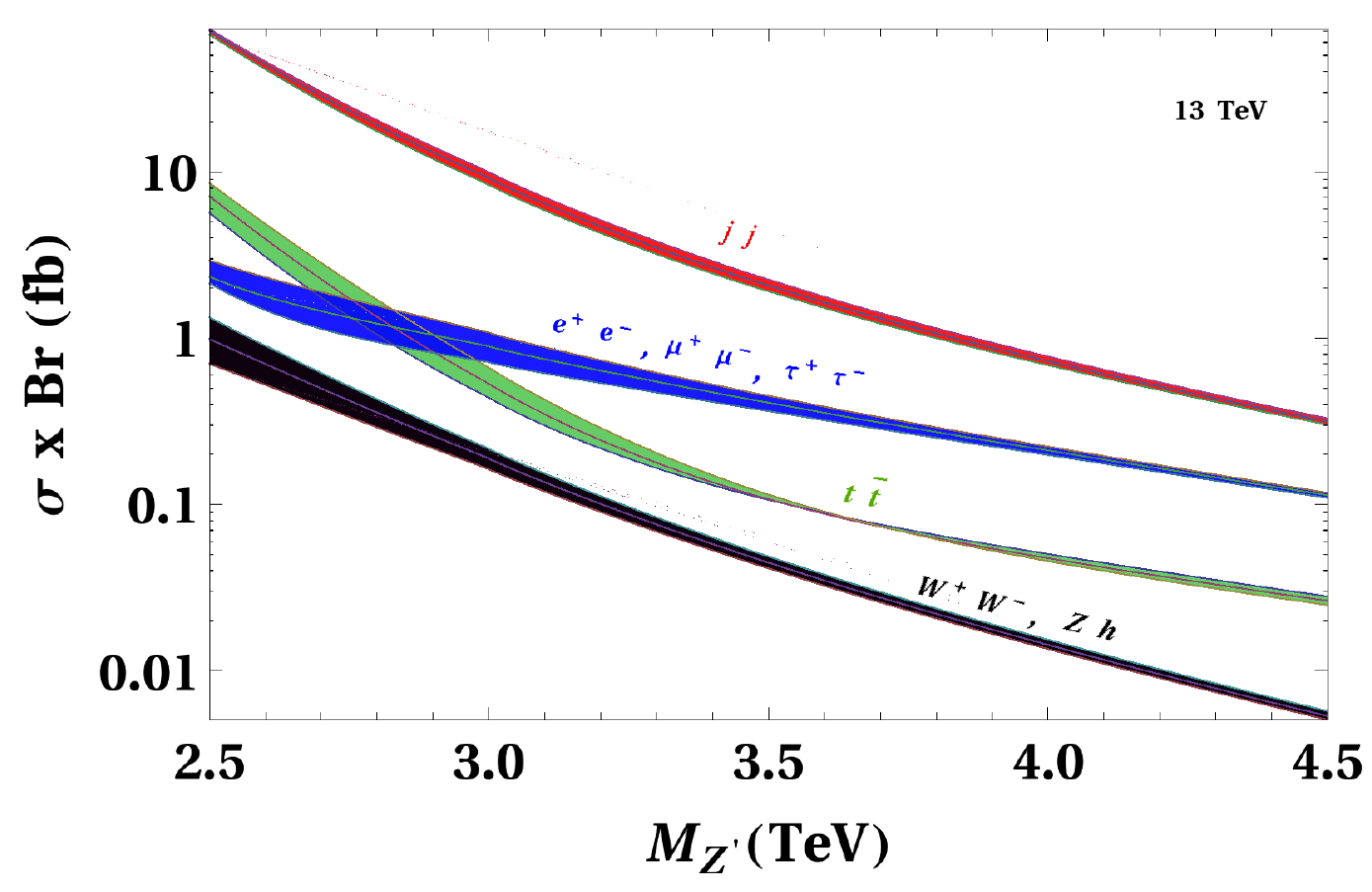}
\caption{\sf Branching fractions (upper plot) and $Z^\prime$--production cross section times branching fractions (lower plot) at the 13 TeV LHC for $M_{W'}=1.8-2$ TeV, with $\gR \approx 0.45-0.6$ and assuming a right handed neutrino mass $m_{N_D}=1.5$ TeV. The coefficients $\gamma_C$ and $\gamma_W$ have been set to zero. All the bands correspond to the mass range $M_{W'}=1.8-2$ TeV (central line in each of them corresponds to $M_{W'}=1.9$ TeV). The $jj$--band is the sum of the partial widths for $\{u\bar{u},\, d\bar{d},\, s\bar{s},\,c\bar{c},\,b\bar{b}\}$, while $\nu\bar{\nu}$--band is the sum of partial widths into SM neutrinos. The bands labelled with several decay modes stand for individual channels.}
\label{Z'-Branching-ratios-cross-sections}
\end{center}
\end{figure}

The branching fractions of the $Z'$ boson for $M_{W'}=1.8-2$ TeV, with $\gR \approx 0.45-0.6$ and assuming a right handed neutrino mass\footnote{The Majorana masses $m_{N^e_R}$ and $m_{N^\tau_R}$ turns out to be equal as the $N^e_R$ and $N^\tau_R$--fields form a Dirac fermion (see~\cite{Dobrescu:2015jvn} for more details).} $m_{N^e_R}=m_{N^\tau_R}=m_{N_D} =1.5$ TeV, has been computed for the fermionic decay channels  $Z^\prime \to \{\nu_L \bar\nu_L,\,N_D\bar N_D,\,\ell^+\ell^-,\,t\bar t,\, jj\}$, and for the gauge-scalar modes $Z^\prime \to \{W^+W^-,\, Z h\}$ in Fig.~\ref{Z'-Branching-ratios-cross-sections} (upper plot). 
The $Z^\prime$--production cross section times branching fractions are computed at the 13 TeV LHC and are displayed in Fig.~\ref{Z'-Branching-ratios-cross-sections} (lower plot). The coefficients $\gamma_C$ and $\gamma_W$ have been set to zero. All the bands in both plots correspond to the mass range $M_{W'}=1.8-2$ TeV (central line in each of them corresponds to $M_{W'}=1.9$ TeV). Fig~\ref{Z'-Branching-ratios-cross-sections} shows a preferred dijet decay channel rather than the top and lepton pair final states respectively. We predict for $M_{W'}=\!1.9\,\text{TeV}$
\begin{itemize}

\item $Z^\prime$-production cross sections of $\{2.3,\,7.1,\,70.2\}\,\text{fb}$ at $M_{Z'}=\!2.5\,\text{TeV}$, through the lepton--pair, top--pair and dijet channels $Z^\prime \to \{\ell^+\ell^-,\,t\bar t,\, jj\}$ respectively, while $0.98\,\text{fb}$ for the gauge--scalar modes $Z^\prime \to \{W^+W^-,\, Z h\}$. The total $Z'$--production cross section of $81.7\,\text{fb}$ at $M_{Z'}=\!2.5\,\text{TeV}$ respectively, mainly dominated by the dijet channel ($86\%$) with complementary small contributions from the top--pair mode ($8.7\%$) and lepton--pair channel ($2.8\%$), plus the $W$--pair and $Z h$ modes ($1.2\%$ both). 

\item At $M_{Z'}=\!4\,\text{TeV}$, the cross sections of $\{0.2,\,0.04,\,0.73\}\,\text{fb}$ for fermionic decay modes correspondingly, and $0.01\,\text{fb}$ for gauge--scalar modes. The total $Z'$--production cross sections of $\sim\!1.0\,\text{fb}$ at $M_{Z'}=\!4\,\text{TeV}$, is dominated mainly by the dijet channel ($71.7\%$) with complementary small contributions from the top--pair mode ($4.6\%$) and lepton--pair channel ($20.7\%$), plus the $W$--pair and $Z h$ modes ($1.4\%$ both). 

\end{itemize}

\begin{figure}
\begin{center}
\includegraphics[scale=0.59]{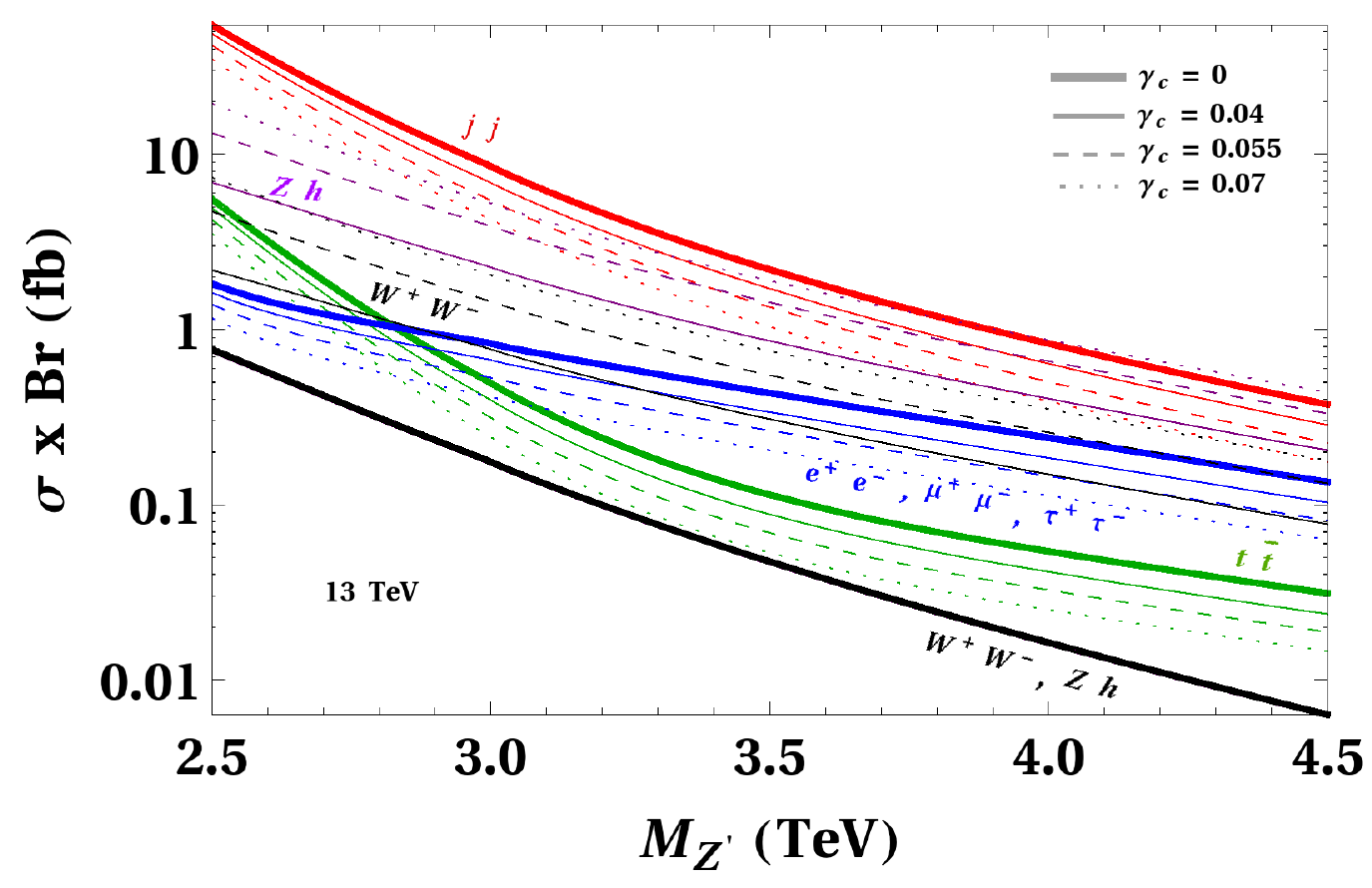}
\\
\vspace*{1cm}
\includegraphics[scale=0.58]{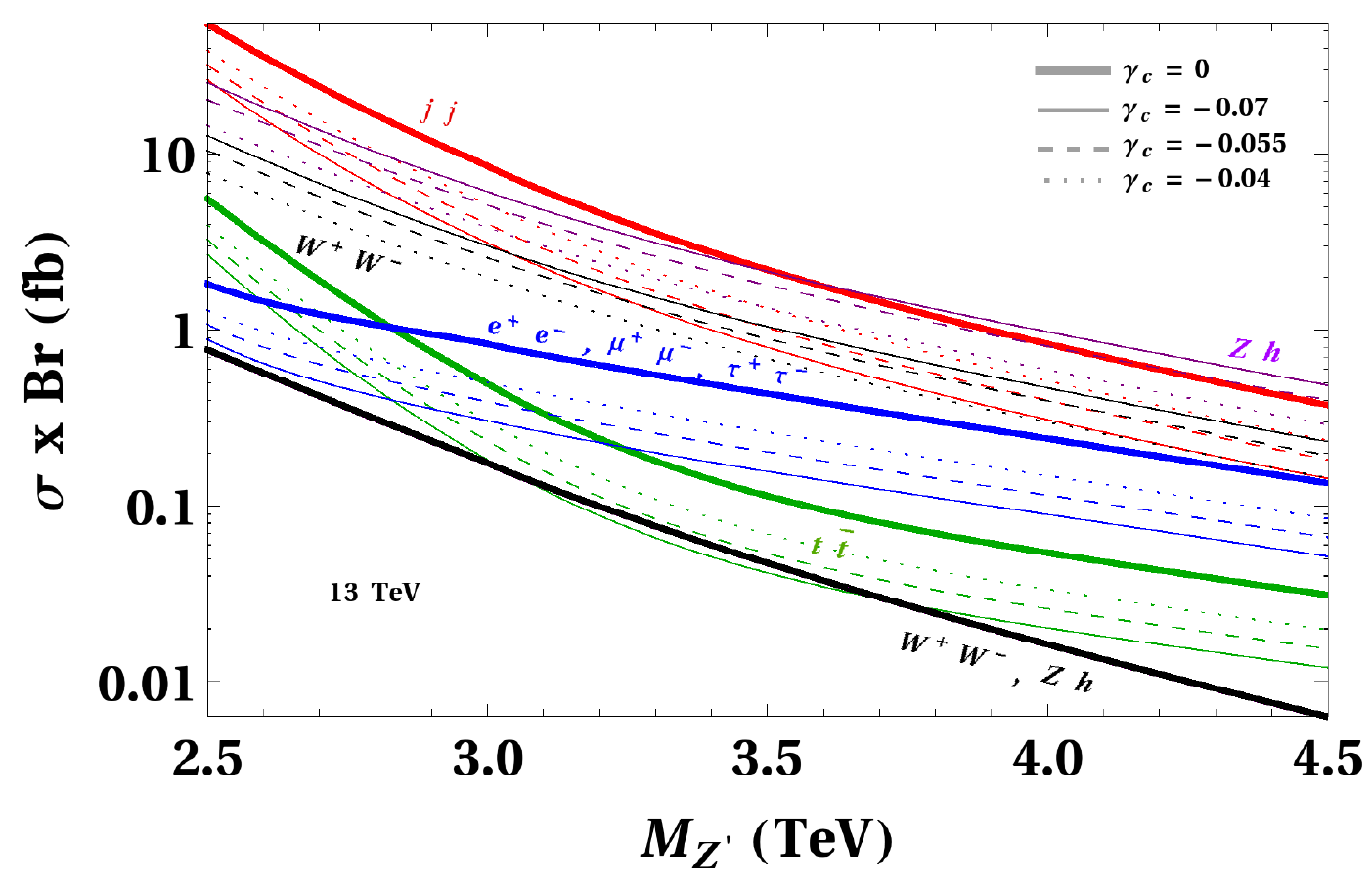}
\caption{\sf $Z^\prime$--production cross section times branching fractions at the 13 TeV LHC for $M_{W'}=1.9$ TeV, and $\gamma_C$ running over the ranges  $[-0.07,\,-0.04]$ (upper plot) and $[0.04,\,0.07]$ (lower plot) following the values in Table~\ref{CLR-CWLR-bounds-W'WZ-W'Wh} and the allowed parameter space for $\gamma_W=0$ and  $\sigma_{jj}(W') \sim 200\,\text{fb}$ (left and right red charts in Fig.~\ref{Coefficients-parameter-space}).
Thick curves correspond to $\gamma_C=0$, whilst the line, dashed and dotted curves stand for the lower, intermediate and upper $\gamma_C$--values according to the allowed ranges.}
\label{Z'-cross-sections-operator-coefficients}
\end{center}
\end{figure}

\nt As it was pointed out before, and according to the couplings in Table~\ref{Z'-fermion-couplings}, the fermionic decay channels are slightly modified by the modifications induced by the operators $\{\cP_{C,LR}(h),\,\cP_{T,LR}(h)\}$ as the involved effective couplings are suppressed by $M_W/M_{Z'}$. Nonetheless, sizeable contributions are triggered on the gauge and gauge-Higgs decay modes once the effective operators are switched on (Table~\ref{Cubic-gauge-left-right-couplings-Z'}). Fig.~\ref{Z'-cross-sections-operator-coefficients} shows the induced effects on the $Z^\prime$-production cross sections for a vanishing operators $\{\cP_{T,LR}(h),\,\cP_{W,LR}(h)\}$ but $\cP_{C,LR}(h)$, at the 13 TeV LHC for $M_{W'}=1.9$ TeV. In particular, the corresponding coefficient $\gamma_C$ runs over the allowed parameter space in Fig.~\ref{Coefficients-parameter-space} for $\gamma_W=0$ and at  $\sigma_{jj}(W') \sim 200\,\text{fb}$ (left and right red charts), i.e, $\gamma_C$ running over the ranges  $[-0.07,\,-0.04]$ (upper plot) and $[0.04,\,0.07]$ (lower plot) from Table~\ref{CLR-CWLR-bounds-W'WZ-W'Wh}. We predict then 
\begin{itemize}

\item In the negative range $\gamma_C=[-0.07,\,-0.04]$, a total $Z'$--production cross sections of 68.1--66.2\,fb at $M_{Z'}=\!2.5,\,\text{TeV}$ and 1.86--1.58\,fb at $M_{Z'}=\!4\,\text{TeV}$. There is an enhancement of (18.8--10.9)\% and (38.7--21.5)\% in the $W$--pair and $Z h$ modes respectively at $M_{Z'}=\!2.5,\,\text{TeV}$, while a raise of (38.9--24.4)\% and (82.6--49.7)\% correspondingly at $M_{Z'}=\!4,\,\text{TeV}$. This leads to an associated enhancement in the total $Z'$--production cross sections of (6.9--3.9)\% at $M_{Z'}=\!2.5,\,\text{TeV}$ and (60.4--36.8)\% at $M_{Z'}=\!4,\,\text{TeV}$ with respect to the vanishing operator case (thick lines in Fig.~\ref{Z'-cross-sections-operator-coefficients} upper plot).

\item In the positive range $\gamma_C=[0.04,\,0.07]$, a total $Z'$--production cross sections of 64.6--66.7\,fb at $M_{Z'}=\!2.5,\,\text{TeV}$ and 1.86--1.58\,fb at $M_{Z'}=\!4\,\text{TeV}$. An enhancement of (2.2--10.3)\% and (9.6--29.5)\% in the $W$--pair and $Z h$ modes respectively at $M_{Z'}=\!2.5,\,\text{TeV}$, while a raise of (11.4--28.9)\% and (33.4--73.7)\% correspondingly at $M_{Z'}=\!4,\,\text{TeV}$. Consequently, an enhancement is observed in the total $Z'$--production cross sections of (1.4--4.7)\% at $M_{Z'}=\!2.5,\,\text{TeV}$ and (22.2--51)\% at $M_{Z'}=\!4\,\text{TeV}$ with respect to the vanishing operator case (thick lines in Fig.~\ref{Z'-cross-sections-operator-coefficients} lower plot).

\end{itemize}

\nt Small deviations from the Goldstone equivalence theorem in the decay widths $\Gamma\left(Z'\to W^+W^-\right)$ and $\Gamma\left(Z'\to Z h\right)$ are induced by the non-zero contributions of the effective operators $\{\cP_{C,LR}(h),\,\cP_{W,LR}(h)\}$. In addition, sizeable enhancement is triggered in those channels due to the effective operators contribution. Such departures become negligible for small coefficients $\gamma_C$ and $\gamma_W$, whose ranges are determined by the $WZ$ and $W h$ excesses in the $W'$--decays studied in Sect.~\ref{WZ-Wh-excess} (Table~\ref{CLR-CWLR-bounds-W'WZ-W'Wh} and Fig.~\ref{Coefficients-parameter-space}). The effective coefficients $a_i$ from the Higgs sector introduced in the $\cF(h)$--definition of~\eqref{F}, in particular $a_{C,LR}$ and $a_{W,LR}$ ,will fix the allowed parameter space $\left(\gamma_C,\,\gamma_W\right)$. Larger values $a_{C,LR},\,a_{W,LR}\sim 1$ will reduce (enhance) the allowed positive (negative) $\gamma_W$--ranges by one order of magnitude, whereas part of the $\gamma_C$--ranges can reach smaller values close to zero for small values of $\gamma_W$. This feature would favour  coefficients $a_{C,LR}$ and $a_{W,LR}$ of order 1 in case of observing 
tiny departures with respect to the cross sections for the gauge--scalar modes $Z^\prime \to \{W^+W^-,\, Z h\}$ in Fig.~\ref{Z'-Branching-ratios-cross-sections}. Sizeable deviations, specially for a larger $M_{Z'}$--values, would point towards intermediate values $a_{C,LR}\sim a_{W,LR}\sim 1/2$ (as shown in Fig.~\ref{Z'-cross-sections-operator-coefficients}) or smaller ones.


\section{Conclusions}
\label{Conclusions}

\nt The small mass peaks observed at ATLAS and CMS near the 1.8-2 TeV is described here via a $W'$--model inspired by the larger local group $\cG=SU(2)_L\times SU(2)_R\times U(1)_{B-L}$ in a non--linear EW dynamical Higgs scenario. The $W'$--production cross section at the 13 TeV LHC is around 700--1200 fb. We analysed the $W'$--production and the constraints on the parameter space of our scenario entailed by the reported excesses in the $WZ$ and $W h$--final states (Table~\ref{CLR-CWLR-bounds-W'WZ-W'Wh} and Fig.~\ref{Coefficients-parameter-space}). We predict the existence of a heavy gauge boson $Z'$ in the 2.5--4 TeV mass range
as well as some of its decay channels testable in the LHC Run II. We determine the cross section times branching fractions, shown in Fig.~\ref{Z'-Branching-ratios-cross-sections}, for the dijet, dilepton and top--pair $Z'$--decay channels at the 13 TeV LHC around 2.3, 7.1, 70.2 fb respectively, for $M_{Z'}= 2.5$ TeV, while one/two orders of magnitude smaller for the dijet/dilepton and top--pair modes at $M_{Z'}= 4$ TeV. Non-zero contributions from the effective operators, and the underlying Higgs sector of the model, will induce sizeable enhancement in the $W^+W^-$ and $Z h$--final states that could be probed in the future LHC Run II.

\section*{Acknowledgements}

\nt The authors of this work acknowledge valuable comments from J.~Gonzalez-Fraile. J.~Y. also acknowledges KITPC financial support during the completion of this work.


\section{$W'$ heavy boson decay widths}
\label{App:W'-Decay-widths}

\nt From the Lagrangian $\LL_{ud W'}$ in~\eqref{W'-charged-currents}, one has 
\be
\Gamma\left(W' \to u\,\bar d\,\right)=\frac{\mathit{g}^2_{\text{\textit{$u_L d_LW'$}}}\,+\,\mathit{g}^2_{\text{\textit{$u_R d_RW'$}}}}{16\pi}\,M_{W'}\,.
\label{W'-ud-width}
\ee

\nt This decay width also applies for the final state $c\,\bar s$, while for $t\,\bar b$ one has
\be
\Gamma\left(W' \to t\,\bar b\,\right)=\frac{\mathit{g}^2_{\text{\textit{$t_L b_LW'$}}}\,+\,\mathit{g}^2_{\text{\textit{$t_R b_RW'$}}}}{16 \pi } \left(1-\frac{3}{2}\,\frac{m^2_t}{M^2_{W'}}\right)\,M_{W'}
\label{W'-tb-width}
\ee

\nt The involve couplings above are given by the corresponding ones in~\eqref{W'-charged-currents} as  
\be
\mathit{g}_{\text{\textit{$u_L d_LW'$}}}=g_L\,\sqrt{\lambda }\,\gamma_C\simeq g_L\,\frac{M_W}{M_{W'}}\,\gamma_C,\,\qquad
\mathit{g}_{\text{\textit{$u_R d_RW'$}}}= - \gR
\label{Charged-couplings-W'}
\ee

\nt Extending the Lagrangian $\LL_{ud W'}$ to the lepton--$W'$ interactions, one has  
\be
\Gamma\left(W' \to \nu_l\,\bar l\,\right)=\frac{\mathit{g}^2_{\text{\textit{$\nu_l l_LW'$}}}}{48 \pi }\,M_{W'}\,,\quad l=e,\,\mu,\,\tau
\label{W'-nue-width}
\ee
\be
\Gamma\left(W' \to N_D\,\bar l\,\right)=\frac{\mathit{g}^2_{\text{\textit{$N_l\,l\,W'$}}}}{48 \pi }\left(1-\frac{3}{2}\frac{M^2_{N_D}}{M^2_{W'}}\right)M_{W'}\,,\quad l=e,\,\tau
\label{W'-Ne-Nmu-width}
\ee

\nt The decay width for the $N^\mu_R\,\bar \mu$--final state is not reported as no $\mu\mu j j$--signal has been oberved so far. The couplings $\mathit{g}_{\text{\textit{$\nu_L l_LW'$}}}$ and $\mathit{g}_{\text{\textit{$N_l\,l_RW'$}}}$ correspond to the couplings in~\eqref{Charged-couplings-W'} respectively. The decay widths for the final states $WZ$ and $W h$ have been given in~\eqref{W'-WZ-width-limiting-case}-\eqref{W'-Wh-width-limiting-case}.


\section{$Z'$ heavy boson decay widths}
\label{App:Z'-Decay-widths}

\nt The $Z'$-heavy boson decays are reported here for the fermionic channels as well as the gauge and gauge--scalar modes. From the effective Lagrangian $\LL_{ffZ'}$ in~\eqref{Z'-neutral-currents} it is possible to compute for the leptonic pair final states
\be
\Gamma\left(Z' \to l^+\,l^-\right)=\frac{\mathit{g}^2_{\text{\textit{$\,l_L l_L Z'$}}}\,+\,\mathit{g}^2_{\text{\textit{$\,l_R l_R Z'$}}}}{24 \pi }\,M_{Z'}\,,\quad l=e,\,\mu,\,\tau
\label{Z'-ee-width}
\ee
\be
\Gamma\left(Z' \to \nu_l\,\bar \nu_l\right)=\frac{\mathit{g}^2_{\text{\textit{$\nu_l \nu_l Z'$}}}}{24 \pi }\, M_{Z'}
\label{Z'-nunu-width}
\ee
\be
\hspace*{-2mm}
\Gamma\left(Z' \to N_D\,\bar N_D\right)=\frac{\mathit{g}^2_{\text{\textit{$N_l N_l Z'$}}}}{24 \pi  }\sqrt{1-4 \frac{M^2_{N_D}}{M_{Z'}^2}} \left(1-\frac{M^2_{N_D}}{M_{Z'}^2}\right)M_{Z'}
\label{Z'-NN-width}
\ee

\nt For the quark--antiquark final states one has
\be
\Gamma\left(Z' \to q\,\bar q\,\right)=\frac{\mathit{g}^2_{\text{\textit{$q_L q_L Z'$}}}\,+\,\mathit{g}^2_{\text{\textit{$q_R q_R Z'$}}}}{8 \pi }\, M_{Z'}\,,\quad q=u,\,d\,.
\label{Z'-ff-width}
\ee

\nt All the involve couplings in~\eqref{Z'-ee-width}-\eqref{Z'-ff-width} $g_{f_L f_L Z'}$ and $g_{f_R f_R Z'}$ with $f=u,d,N,\nu,e$, are listed in Table~\ref{Z'-fermion-couplings}. From the effective Lagrangian $\LL_{W W Z'}$ in~\eqref{Cubic-gauge-left-right-interactions-Z'} one has, for the $W$--pair final state
\be
\Gamma\left(Z'\to W^+W^-\right)=\frac{\left(g_{WWZ'}^{\text{(2)}}\right)^2}{192 \pi }\,\frac{M^4_{Z'}}{M^4_W}\,M_{Z'}\,.
\label{Z'-WW-width}
\ee

\nt The extra factor $M^4_{Z'}/M^4_W$ comes from the longitudinal helicity component in the decay $Z'\to W^+W^-$, being compensated by the quadratic inverse term from $g_{WWZ'}^{\text{(2)}}$ for a vanishing operator contribution (look at Table~\ref{Cubic-gauge-left-right-couplings-Z'}). A non-zero operator contribution leads to additional terms enhanced by the extra factor as it is reflected in Fig~\ref{Z'-cross-sections-operator-coefficients}. Finally, for the $Z h$--final state, one has
\be
\Gamma\left(Z'\to Z h
\right)\simeq \frac{g_{\text{\textit{$h$}}\text{\textit{$ $}}ZZ'}^{\text{(1)}}}{192 \pi\,c^2_W}\,\left( g_{\text{\textit{$h$}}\text{\textit{$ $}}ZZ'}^{\text{(1)}}\,+\,\frac{M^2_Z}{M^2_{Z'}} g_{\text{\textit{$h$}}\text{\textit{$ $}}ZZ'}^{\text{(2)}}\right)\,\frac{M^4_{Z'}}{M^4_Z}\,M_{Z'}\,.
\label{Z'-Zh-width}
\ee

\nt The involve couplings are listed in Table~\ref{Cubic-gauge-left-right-couplings-Z'}.
%

\providecommand{\href}[2]{#2}\begingroup\raggedright\endgroup

\end{document}